\documentclass[12pt]{article}
\usepackage[latin1]{inputenc}
\usepackage[T1]{fontenc}
\usepackage[english]{babel}
\usepackage{epsfig}
\usepackage{amssymb}
\usepackage[centertags]{amsmath}
\usepackage{amsthm}
\usepackage{amsfonts}
\usepackage{makeidx}
\usepackage{graphics}
\usepackage{graphicx}
\usepackage{psfrag}
\usepackage{subfigure}
\usepackage{dsfont}
\usepackage{eucal}
\usepackage{eufrak}
\usepackage{color}
\usepackage{bm}
\newcommand{\bc}{\begin{center}}
\newcommand{\ec}{\end{center}}
\newcommand{\be}{\begin{equation}}
\newcommand{\ee}{\end{equation}}
\newcommand{\bea}{\begin{eqnarray}}
\newcommand{\eea}{\end{eqnarray}}
\newcommand{\ba}{\begin{array}}
\newcommand{\ea}{\end{array}}

\newcommand{\bfg}{\begin{figure}[htbp]}
\newcommand{\efg}{\end{figure}}
\newcommand{\pr}{Phys. Rev. }

\def \de {\partial}

\def \a {\alpha}
\def \b {\beta}
\def \g {\gamma}
\def \G {\Gamma}
\def \d {\delta}

\def \ep {\epsilon}
\def \ve {\varepsilon}

\def \l {\lambda}

\def \m {\mu}
\def \n {\nu}

\def \be {\begin{equation}}
\def \ee {\end{equation}}
\def \bea {\begin{eqnarray}}
\def \eea {\end{eqnarray}}
\def \non {\nonumber}

\def \pr {\prime}

\def\laq{~\raise 0.4ex\hbox{$<$}\kern -0.8em\lower 0.62
ex\hbox{$\sim$}~}
\def\gaq{~\raise 0.4ex\hbox{$>$}\kern -0.7em\lower 0.62
ex\hbox{$\sim$}~}

\begin{document}

\vspace{0.5 cm}
\begin{center}
{\large \textbf{Towards a consistent $AdS$/QCD dictionary}}
\vspace{1.cm}

Frédéric Jugeau$^{1,2}$\\
\vspace{0.2 cm}
\textit{$^1$ Institute of High Energy Physics, Chinese Academy of Sciences,\\
P. O. Box 918(4), Beijing 100049, China\\
$^2$ Theoretical Physics Center for Science Facilities, Chinese
Academy of Sciences,\\
Beijing 100049, China}\\
\vspace{0.3 cm}
jugeau@ihep.ac.cn\\
\vspace{0.2 cm} (Dated: February 12, 2009)
\end{center}
\par
\renewcommand{\thefootnote}{\fnsymbol{footnote}}
\vspace{0.5 cm}

\begin{center}
{\large Abstract}
\end{center}
This note focuses on the large$-N$ behaviour of the Hard Wall
model of QCD and clarifies the $AdS$/QCD dictionary formulated on
the basis of the $AdS$/CFT correspondence. It is shown how
short-distance studies performed in the framework of the $AdS$/QCD
Soft Wall model allow one to determine unambiguously the chiral
symmetry breaking function in the Hard Wall model. Especially, the
different forms of the field/operator prescription are emphasized.
The large$-N$ behaviour of the Hard Wall model is then checked
considering form factors of the pion in the chiral limit.

\par
\vspace{1.0 cm} PACS numbers: 11.15.Pg, 11.25.Tq,
11.30.Rd,14.40.Aq.
\hspace*{2.65 cm}pion form factors.
\newpage

\tableofcontents

\section{Introduction}

Although considerable simplification is obtained in the large$-N$
limit of QCD, corresponding to the planar diagram approximation of
the theory \cite{Hooft}, its resolution in the \emph{infrared}
(IR) regime remains one of the most intricate long-standing
problem for particle physicists. For instance, it has not been
possible, up to now, to derive an analytical expression for the
interaction force between the two constituents of a quarkonium
$q\overline{q}'$ from QCD itself. Nevertheless, a new hope arose
when J.M. Maldacena conjectured in 1998 that a four-dimensional
strongly-coupled gauge theory could be described equivalently by a
weakly-coupled string theory living in a higher-dimensional
space-time \cite{Maldacena AdS/CFT}. Strictly speaking, this
so-called $AdS$/CFT correspondence deals, on the one hand, with a
four-dimensional supersymmetric and conformal strongly-coupled
$SU(N)$ Yang-Mills theory in the large$-N$ 't Hooft limit and, on
the other hand, with a weakly-coupled type-IIB (oriented closed)
superstring theory in the low-energy or supergravity limit (where
the stringy effects $O({\ell_s}^2)\sim O(\frac{1}{\sqrt{N}})$ with
$\ell_s$ the typical string length are neglected at first
approximation). The bulk in which propagate the non-interacting
strings is the product of a five-dimensional anti-de Sitter
space-time $AdS_5$ with a 5-sphere $S^5$. As for the dual gauge
field theory, it is defined on the four-dimensional $AdS_5$
boundary (for a detailed explanation of the correspondence, the
reader is referred to, e.g., \cite{correspondance explanation}).

It is useless to say that its applicability to a confining gauge
theory is far to be obvious. Indeed, as a theory of the Standard
Model, QCD contains no supersymmetric partner of the quarks and of
the gluons. Moreover, the asymptotic freedom of QCD introduces a
mass scale $\Lambda_{QCD}$ at which arises the confinement
mechanism responsible of the cohesion of the quarks within the
hadrons. Nonetheless, it could be possible to modify Maldacena's
conjecture in order to treat such theories as well. Some
indications are indeed stimulating which conduce to pursue in this
direction: for asymptotically high energies, QCD presents a nearly
conformal behavior since the running strong coupling constant
vanishes logarithmically and the quark masses are negligible in
comparison with the energies brought into play in the
\emph{ultraviolet} (UV) regime. On the other hand, at lower
energies, the QCD $\beta$ function could have an IR fixed point
related to the 'freezing behaviour' of the coupling
\cite{vanishing beta function} and thus to a conformal window for
QCD \cite{conformal window}. Two complementary approaches exist on
the market which aim at identifying the dual theory of QCD. In the
historically first approach, the so-called \emph{top-down}
approach \cite{Witten conformal breaking}, the conformal
invariance and the supersymmetry are broken, respectively, by
compactification (the compactification radius giving rise to a
dimensionful parameter, namely, the mass gap of QCD) and
appropriate boundary conditions on the compactified dimensions.
The $AdS$ geometry is then deformed into a Schwarzschild black
hole$-AdS$ geometry where the event horizon plays the role of an
IR brane whose the location is related to the Beckenstein-Hawking
temperature (\emph{i.e.} the thermal temperature of the gauge
theory). On the other hand, according to the \emph{bottom-up} or
AdS/QCD approach, one focuses on the phenomenological properties
of QCD and uses them as guidelines in order to construct its
five-dimensional dual theory.

The explicit carrying out of the holographic principle
\cite{Witten} consists in associating, in a one-to-one
correspondence, a dual $p$-form bulk field, generically denoted
$\phi(x,z)$ where $x^\mu$ ($\mu=0,\ldots,3$) are the usual $4d$
Minkowski coordinates and $z$ is the $5^{th}$ holographic
coordinate (dual to the energy scale at which is observed the
conformal field theory (CFT) defined on the $AdS$ boundary at
$z\rightarrow0$), with a $p$-form CFT operator $\mathcal{O}(x)$
with conformal dimension $\Delta$. The square mass $m_{AdS}^2$ of
the bulk field is given in terms of $\Delta$ and of the rank $p$
of the form according to the $AdS$/CFT relation:
\begin{equation}\label{AdS/CFT mass relation}
m_{AdS}^2R^2=(\Delta-p)(\Delta+p-d)
\end{equation}
where $d=4$ is the dimension of the boundary space-time and $R$ is
the so-called $AdS$ radius (having the dimension of a length). The
relevance of the $AdS$/CFT correspondence comes from the fact that
it allows one to derive any $n$-point correlation functions of the
boundary conformal field theory by means of its dual $AdS$ theory
\cite{conformal correlator}. The method is based on the
equivalence between the generating functional of the connected
correlators of the $4d$ CFT and the $AdS_5$ partition function.
The conjecture is precisely that \cite{Witten}
\begin{equation}\label{generating}
\biggl\langle e^{i\int_{\de AdS_5} d^4
x\;\mathcal{O}(x)\,\phi_0(x)}\biggr\rangle_{CFT}=e^{iS_{5d}[\phi(x,z)]}\biggl|_{\phi\underset{z\to0}{\to}\phi_0}
\end{equation}
where $S_{5d}[\phi(x,z)]$ is the supergravity action (obtained by
inserting the field solutions of the classical equations of
motion, hence the absence of path integral on the $\phi$ field
configurations on the \emph{r.h.s.} of \eqref{generating}). On the
other hand, the bulk field $\phi(x,z)$ is required to approach
asymptotically the source field $\phi_0(x)$ associated with the
operator $\mathcal{O}(x)$ defined on the boundary. As for a
massive scalar bulk field (\emph{i.e.} $p=0$), its near-boundary
behaviour when $z\rightarrow0$ turns out to be
\begin{equation}\label{near-boundary behaviour}
\phi(x,z)\underset{z\rightarrow0}{\rightarrow}z^{d-\Delta}\Big[\phi_0(x)+O(z^2)\Big]+z^{\Delta}\Big[A(x)+O(z^2)\Big]
\end{equation}
where $A(x)=\frac{\langle\mathcal{O}(x)\rangle}{2d-\Delta}$ is
argued to be related to the vacuum expectation value of the CFT
operator \cite{KW}. In general, the relation between $\phi(x,z)$
and $\phi_0(x)$ involves the so-called bulk-to-boundary propagator
$\phi(x-x';z,0)$:
\begin{equation}\label{convolution}
\phi(x,z)=\int d^d x^{\pr}\,\phi(x-x^\prime;z,0)\,\phi_0(x^\prime)
\end{equation}
and the derivation of the correlation functions from supergravity
side gives expressions involving such propagators. In the $4d$
Fourier space with respect to the boundary coordinates $x^\mu$,
the convolution \eqref{convolution} reads
\begin{equation}\label{fourier convolution}
\tilde{\phi}(q,z)=\tilde{\phi}(q^2,z)\tilde{\phi}_0(q)\;\;,
\end{equation}
written in terms of Fourier transformed fields defined by
\begin{equation}\label{fourier convolution2}
\phi(x,z)=\int \frac{d^dx}{(2\pi)^d}e^{iq\cdot
x}\tilde{\phi}(q,z)\;\;.
\end{equation}

Any attempt to establish the $AdS$/QCD correspondence involving a
non-conformal and non-supersymmetric Yang-Mills theory, if there
is such a duality, should use a process of successive
modifications, keeping in touch as much as possible with the
checked $AdS$/CFT prescription briefly described above. Several
scenarios have been proposed which break the conformal invariance
by warping the $AdS_5$ geometry in the IR region at large $z$. In
the framework of the \emph{bottom-up} approach, the Hard Wall
model \cite{PT,EKSS,pomarol} consists in truncating $AdS_5$ by
compactifying the holographic coordinate $0\leq z\leq z_m$ where
the IR cutoff is related to the QCD mass gap
$z_m\simeq1/\Lambda_{QCD}$. Thus, QCD is crudely seen as a
conformal field theory up to a specific energy scale at which the
conformal symmetry is sharply broken. The light hadron
spectroscopy \cite{hard wall model spectrum}, the meson and
nucleon form factors \cite{hard wall model form
factors,GR,carlson}, the two-point correlation functions
\cite{hard wall model correlator}, the deep inelastic scattering
structure functions \cite{hard wall model DIS}, the chiral
symmetry breaking mechanism and the $U(1)$ problem \cite{hard wall
model chiral} have been investigated. In particular, it has been
shown that the spectrum satisfies a Kaluza-Klein behavior
$m_n^2\sim n^2$ instead of the expected Regge behavior $m_n^2\sim
n$ ($n$ is the radial excitation number). To remedy this
shortcoming, the Soft Wall model has been proposed which
introduces a smooth IR cutoff through a background dilaton field
\cite{KKSS,soft wall model,soft wall scalar}. The hadrons have
also been studied in the framework of D$p$/D$q$-brane systems
\cite{huang}, as in the Sakai-Sugimoto
D4/D8-$\overline{\textrm{D8}}$ model \cite{SS}. The
$(p+1)$-dimensional SU(N) QCD-like theory is described by a stack
of $N$ coincident D$p$-branes whereas the flavour degrees of
freedom are introduced by $n_f$ D$q$-branes in the probe limit
where their back-reaction can be neglected \cite{KK}.

The five-dimensional holographic model proposed in \cite{EKSS}
focuses on the low-energy phenomenology of the lightest mesons.
Being interested in the chiral dynamics of QCD, only a small
number of operators are relevant. In particular, in order to
account for the chiral symmetry breaking, a scalar bulk field
$v(z)$, dual to the vacuum expectation value of the operator
$\overline{q}q$, is introduced in the $5d$ effective action.
According to \eqref{near-boundary behaviour} with $d=4$ and
$\Delta=3$, it behaves near the UV brane as
\begin{equation}\label{UV boundary condition of v}
\frac{1}{z}v(z)\underset{z\to0}{\to}\frac{m_q}{R}
\end{equation}
where the source $m_q$ is the light quark mass. It is not
surprising to find the $AdS$ radius $R$ on the \emph{r.h.s.} of
\eqref{UV boundary condition of v} as it is the only dimensionful
parameter associated with the UV region of the holographic space
(the hard cutoff $z_m$ being related to the IR region). As a
matter of fact, the exact solution of the linearized equation of
motion (see \eqref{HW vev EOM} and \eqref{v} below) is:
\begin{equation}
v(z)=\frac{m_q}{R}z+\frac{\overline{\sigma}}{R}z^3
\end{equation}
where $\overline{\sigma}$ is rather systematically identified in
the literature with the chiral condensate
$\overline{\sigma}=-\langle\overline{q}q\rangle$. As pointed out
in \cite{CCW}, such a naive identification leads to
inconsistencies regarding the large$-N$ behaviour since $m_q\sim
O(N^0)$ and $\langle\overline{q}q\rangle\sim O(N)$ \cite{Pich}.
The aim of this note is to reexamine the large$-N$ behaviour of
the Hard Wall $AdS$/QCD model by considering the static and
dynamic properties of the pion in the chiral limit. Especially, it
is emphasized that no systematic modification of the standard
field/operator correspondence is required in order to have a
consistent large-$N$ behaviour in the Hard Wall model (in other
words, we shall still follow the UV boundary condition \eqref{UV
boundary condition of v}). The letter is organized as follows:
after reviewing in section 2 the Hard Wall model in details, we
revise the derivation of the Gell-Mann-Oakes-Renner relation
(GMOR) in section 3, that will allow us to identify unambiguously
the quark mass and the chiral condensate entering the expression
of the so-called chiral symmetry breaking function $v(z)$. These
studies shed light on the precise formulation of the $AdS$/QCD
dictionary. Finally, we check the large$-N$ behaviour of the Hard
Wall model by considering the pion electromagnetic and
gravitational form factors from which can be derived mean square
radii and coupling constants.

\section{The Hard Wall model for a $5d$ gravity dual of QCD}

Following \cite{EKSS}, we consider a $5d$ conformally flat
space-time (the bulk) described by the metric ($M,N=0,\ldots,4$
and $\m,\n=0,\ldots,3$):
\begin{equation}\label{IR hard wall metric}
ds^2=g_{MN}(z)dx^Mdx^N=\frac{R^2}{z^2}\Big(\eta_{\m\n}dx^{\m}dx^{\n}-dz^2\Big)
\end{equation}
with
\begin{equation}
g_{MN}(z)=\frac{R^2}{z^2}\eta_{MN}
\end{equation}
and
\begin{equation}\label{5d flat metric}
\eta_{MN}= \left(
\begin{array}{ll}
\eta_{\m\n}&0\\
0&-1
\end{array}
\right)
\end{equation}
where $\eta_{\m\n}=\textrm{diag}(+1,-1,-1,-1)$ is the flat metric
tensor of the $4d$ Minkowski boundary space. The holographic
coordinate $z$ runs from zero to the hard wall located at $z_m$.
The $5d$ effective action able to reproduce the chiral properties
of QCD is:
\begin{eqnarray}
S_{5d}&=&S_{5d}^{(grav.)}+S_{5d}^{(matter)}\label{action}\\
S_{5d}&=&-\frac{1}{2\kappa^2}\int
d^5x\sqrt{g}\,\Big(\mathcal{R}+\Lambda\Big)\label{gravity part}\\
&&+\frac{1}{k}\int d^5
x\sqrt{g}\,Tr\,\Big\{|DX|^2-m_5^2\,X^2-\frac{1}{4g_5^2}(G_L^2+G_R^2)\Big\}\label{IR
hard wall action}
\end{eqnarray}
where the matter contribution $S_{5d}^{(matter)}$ governing the
dynamics of the bulk fields is supplemented with the
five-dimensional gravity action $S_{5d}^{(grav.)}$
\cite{carlson,csaki}. $g=\textrm{det}(g_{MN})$ is the determinant
of the tensor metric, $\kappa^2$ is the $5d$ Newton constant and
$\mathcal{R}$ is the scalar curvature of the $AdS_5$ space-time,
inversely related to the $AdS$ radius $R$. With our mostly minus
signature \eqref{5d flat metric}, the cosmological constant of the
$AdS_5$ space-time is negative
$\Lambda=-\frac{3}{5}\mathcal{R}=-\frac{12}{R^2}$. The trace
operator $Tr$ acts on the chiral flavour internal space since,
according to the AdS/CFT dictionary, the chiral symmetry
$SU(n_F)_L\times SU(n_F)_R$ at the boundary must be gauged in the
bulk. In the sequel, we will consider the case of two light
flavours $n_F=2$. The overall parameter $k$ has the dimension of a
length while the gauge coupling $g^2_5$ is dimensionless (which is
also the convention used by \cite{soft wall scalar} in the
framework of the Soft Wall model).

In the chiral limit that we will consider here, the relevant QCD
operators are the left-handed and the right-handed currents,
$j_{L\,\m}^a=\overline{q}_L\g_{\m}t^aq_L$ and
$j_{R\,\m}^a=\overline{q}_R\g_{\m}t^aq_R$, corresponding to the
$SU(2)_L\times SU(2)_R$ chiral symmetry group and the chiral order
parameter $\overline{q}_R^iq_L^j$ (when $n_F=2$,
$t^a=\frac{\sigma^a}{2}$ are the Pauli matrices with $a=1,2,3$ and
$i,j=1,2$ the flavour adjoint and fundamental indices,
respectively). All the other operators are supposed to not have
significant role at low energy and can be neglected when writing
the effective action. The bulk field content of the $5d$ dual
theory consists then of the left-handed and right-handed gauge
fields $A_{L_M}^a(x,z)$ and $A_{R_M}^a(x,z)$. As for the operator
$\overline{q}_R^iq_L^j$, it is dual to the bi-fundamental scalar
field:
\begin{equation}\label{scalar bulk field}
X^{ij}(x,z)=\Big(\frac{v(z)}{2}+S(x,z)\Big)^{ik}\Big(e^{2i\pi^b(x,z)t^b}\Big)^{kj}
\end{equation}
which is massive $m_{5}^2R^2=-3$ according to \eqref{AdS/CFT mass
relation} with $p=0$ and $\Delta=3$ and involves the bulk fields
$S^{ik}(x,z)$ dual to the scalar mesons, the dimensionless
pseudo-scalar fields $\pi^b(x,z)$ and the background field
$v^{ik}(z)$, written as $v^{ik}(z)=v(z)\d^{ik}$ in the isospin
limit, which describes the explicit ($m_q\neq0$) and implicit
($\langle\overline{q}q\rangle\neq0$) chiral symmetry breaking.

The non-Abelian field strength tensors
\begin{equation}
G_{{L,R}_{MN}}=\de_MA_{{L,R}_N}-\de_N
A_{{L,R}_M}-i[A_{{L,R}_M},A_{{L,R}_N}]
\end{equation}
and the covariant derivative
\begin{equation}\label{covariant derivative scalar field}
D_MX=\partial_MX-iA_{L\,M}X+iXA_{R\,M}
\end{equation}
can be expressed in \eqref{IR hard wall action} in terms of the
vector $V_M^a=(A^a_L+A^a_R)/2$ and axial-vector
$A_M^a=(A^a_L-A^a_R)/2$ bulk fields dual to the currents
$j^a_{V\,\m}=\overline{q}\g_{\m}t^aq$ and
$j^a_{A\,\m}=\overline{q}\g_{\m}\g_5t^aq$ respectively. One
obtains $G_L^2+G_R^2=2(G_V^2+G_A^2)$ and
$D_MX=\de_MX-i[V_M,X]-i\{A_M,X\}$. Moreover, under the parity, the
transformation rules of $G_{V_{MN}}$ and $G_{A_{MN}}$ are opposite
which enables us to identify:
\begin{eqnarray}
G_{V_{MN}}&=&F_{V_{MN}}-i[V_M,V_N]-i[A_M,A_N]\;\;,\label{vector non abelian strengh tensor}\\
G_{A_{MN}}&=&F_{A_{MN}}-i[V_M,A_N]-i[A_M,V_N]\label{axial non
abelian strengh tensor}
\end{eqnarray}
with $F_{V_{MN}}\equiv\partial_{M}V_N-\partial_NV_M$ and
$F_{A_{MN}}=\partial_MA_N-\partial_NA_M$ the Abelian vector and
axial-vector field strength tensors. The quadratic part of the
$5d$ effective action in \eqref{IR hard wall action} takes then
the form (we shall not consider the scalar bulk field $S(x,z)$ in
the following):
\begin{equation}\label{quadratic action}
S_{5d}^{(2)}=\frac{1}{k}\int d^5
x\sqrt{g}\,Tr\,\Big\{\frac{1}{4}\Big(g^{MN}\de_Mv\de_Nv-m_5^2v^2\Big)+v^2g^{MN}(\de_M\pi-A_M)(\de_N\pi-A_N)-\frac{1}{2g_5^2}(F_A^2+F_V^2)\Big\}
\end{equation}
from which can be derived the linearized equations of motion for
the different bulk fields. It is straightforward to write the
equations satisfied by the chiral symmetry breaking function
$v(z)$ and the Fourier transform \eqref{fourier convolution2} of
the vector field $\tilde{V}_{\mu}^a$ (transverse since the vector
current is conserved). We have:
\begin{equation}\label{HW vev EOM}
\de_z\Big(\frac{1}{z^3}\de_zv(z)\Big)+\frac{3}{z^5}v(z)=0
\end{equation}
and (in the axial-like gauge $\tilde{V}^a_z=0$)
\begin{equation}\label{HW vector eq of motion}
\de_z\Big(\frac{1}{z}\de_z\tilde{V}^a_{\m}(q,z)\Big)+\frac{q^2}{z}\tilde{V}^a_{\m}(q,z)=0\;\;.
\end{equation}
On the other hand, provided that $v(z)$ is non-zero, the
longitudinal components of the axial-vector bulk field $A_M^a$,
defined as $\de_M\phi^a=A^a_M-A^a_{\perp\,M}$, are physical
degrees of freedom and are related to the pion while the
(normalizable modes of the) transverse components $A^a_{\perp\,M}$
describe the axial-vector mesons. Since the pion field $\pi^a$ in
\eqref{scalar bulk field} and the longitudinal part $\de\phi^a$ of
the axial-vector field describe both the pseudo-scalar mode
(particle and external current), they must be taken into account
if we want to treat the pion in a consistent way. The pion mode is
thus described by two coupled differential equations. Indeed, the
variation of \eqref{quadratic action} with respect to the
axial-vector field $\tilde{A}_M^a$ (in the axial-like gauge
$\tilde{A}_z^a=0$ and considering the transverse property
$q_{\mu}\tilde{A}^{\mu\,a}_{\perp}=0$ in the $4d$ boundary
space-time) gives:
\begin{equation}\label{HW b-to-b axial-vector EOM}
\de_z\Big(\frac{1}{z}\de_z\tilde{A}^a_{\perp\,\m}(q,z)\Big)+\frac{1}{z}\Big(q^2-\frac{R^2g_5^2v(z)^2}{z^2}\Big)\tilde{A}^a_{\perp\,\m}(q,z)=0\;\;,
\end{equation}
\begin{equation}\label{HW pion EOM 2}
\de_z\Big(\frac{1}{z}\de_z\tilde{\phi}^a(q,z)\Big)-\frac{R^2g_5^2v(z)^2}{z^3}\Big(\tilde{\phi}^a(q,z)-\tilde{\pi}^a(q,z)\Big)=0\;\;,
\end{equation}
\begin{equation}\label{HW pion EOM 1}
q^2\de_z\tilde{\phi}^a(q,z)-\frac{R^2g_5^2v(z)^2}{z^2}\de_z\tilde{\pi}^a(q,z)=0\;\;.
\end{equation}

In terms of the vector and axial-vector bulk-to-boundary
propagators \eqref{fourier convolution}, the linearized equations
of motion \eqref{HW vector eq of motion} and \eqref{HW b-to-b
axial-vector EOM} write out:
\begin{equation}\label{HWM vector b-t-b}
\Big[\de_z\Big(\frac{1}{z}\de_z\Big)+\frac{q^2}{z}\Big]\tilde{V}(q^2,z)=0
\end{equation}
and
\begin{equation}\label{HW axial-vector b-to-b}
\Big[\de_z\Big(\frac{1}{z}\de_z\Big)+\frac{1}{z}\Big(q^2-\frac{R^2g_5^2v(z)^2}{z^2}\Big)\Big]\tilde{A}(q^2,z)=0\;\;.
\end{equation}
It is worth pointing out that these propagators satisfy, when
$v(z)=0$ in \eqref{HW axial-vector b-to-b}, the same equation of
motion. It results a nonphysical degenerate mass spectrum between
the $\rho$ and the $a_1$ mesons, removed once the chiral symmetry
is broken. Hence, the importance of having a consistent expression
for $v(z)$. The propagators satisfy the UV Dirichlet boundary
conditions:
\begin{equation}\label{HWM UV bc}
\tilde{V}(q^2,0)=\tilde{A}(q^2,0)=1
\end{equation}
in order to interpret $\tilde{V}^a_{0\,\m}(q)$ and
$\tilde{A}^a_{0\perp\,\m}(q)$ as source fields and Neumann
boundary conditions at the IR brane ($z=z_m$):
\begin{equation}\label{HWM IR bc}
\de_z\tilde{V}(q^2,z)\Big|_{z=z_m}=\de_z\tilde{A}(q^2,z)\Big|_{z=z_m}=0\;\;.
\end{equation}
For space-like momentum $Q^2=-q^2>0$, the solution of \eqref{HWM
vector b-t-b} is ($Q\equiv\sqrt{Q^2}$):
\begin{equation}\label{HW space-like vector b-to-b}
\tilde{V}(Q^2,z)=Qz\Big(K_1(Qz)+\frac{K_0(Qz_m)}{I_0(Qz_m)}I_1(Qz)\Big)\;\;,
\end{equation}
written in terms of modified Bessel functions of the first and
second kind. Its low-$z$ and low-$Q^2$ expansions are
respectively:
\begin{equation}
\tilde{V}(Q^2,z)\underset{z\to0}{=}1+\frac{Q^2z^2}{4}\Big(\ln(Q^2z^2)-\ln4+2\g_E-1+2\frac{K_0(Qz_m)}{I_0(Qz_m)}\Big)+O(z^4)
\end{equation}
and
\begin{equation}\label{Q2 expansion of vector b-to-b propagator}
\tilde{V}(Q^2,z)\underset{Q^2\to0}{=}1-\frac{Q^2z^2}{4}\big(1-2\ln(\frac{z}{z_m})\big)+O(Q^4)
\end{equation}
which shows that $\tilde{V}(0,z)=1$.

The study of the two-point vector current correlation function in
the framework of the Hard Wall model proceeds from the equivalence
\eqref{generating} applied to QCD. We start from the vector
quadratic part of the action \eqref{quadratic action} which,
evaluated on the solution to the classical $5d$ equation of motion
\eqref{HW vector eq of motion}, gives only the surface term:
\begin{equation}\label{pole correlator AdS}
S_{5d}^{(2)}=-\frac{1}{2}\frac{R}{kg_5^2}\int\frac{d^4q}{(2\pi)^4}\tilde{V}_{0\,\m}^a(q)\Pi^{\m\n}_{\perp}(q)\tilde{V}_{0\,\n}^a(-q)\Big[\tilde{V}(q^2,z)\frac{1}{z}\de_z\tilde{V}(q^2,z)\Big]\Big|_{z=\epsilon\to0}
\end{equation}
with
$\Pi^{\m\n}_{\perp}(q)\equiv\eta^{\m\n}-\frac{q^{\m}q^{\n}}{q^2}$
the orthogonal projector. Deriving twice with respect to the
vector source fields, one obtains \cite{EKSS}:
\begin{equation}\label{HW vector AdS correlation function}
{\Pi^{(AdS)}_{V}}^{ab}_{\m\n}(q)=\d^{ab}(q_{\m}q_{\n}-q^2\eta_{\m\n})\Big(\frac{1}{q^2}\frac{R}{kg_5^2}\tilde{V}(q^2,z)\frac{1}{z}\de_z\tilde{V}(q^2,z)\Big)\Big|_{z=\epsilon\to0}
\end{equation}
which has to be compared with the expression in QCD:
\begin{equation}\label{HW vector QCD correlator}
{\Pi^{(QCD)}_{V}}^{ab}_{\m\n}(q)\equiv i\int d^4x\,e^{iq\cdot
x}\langle0|T[j_{V\m}^a(x)j_{V\n}^b(0)]|0\rangle=\d^{ab}(q_{\m}q_{\n}-q^2\eta_{\m\n})\Pi_V^{(QCD)}(q^2)\;\;.
\end{equation}
In general, the correlation functions derived in $AdS$ side are
singular when $z\to0$ and must be regularized by an UV cutoff
$z_{min}\equiv\epsilon$ identified with the renormalization scale
$1/\n$. The matching \cite{EKSS,Reinders} between the perturbative
parts of \eqref{HW vector AdS correlation function} and \eqref{HW
vector QCD correlator} allows one to extract the factor
$\frac{kg^2_5}{R}$:
\begin{equation}\label{HW overall constant vector meson}
\left.
\begin{array}{l}
\Pi^{(AdS,\,pert.)}_V(Q^2)=-\frac{R}{2kg^2_5}\ln(Q^2z^2)\Big|_{z=\frac{1}{\n}}\\
\Pi^{(QCD,\,pert.)}_V(Q^2)=-\frac{N}{24\pi^2}\ln\Big(\frac{Q^2}{\n^2}\Big)
\end{array}
\right\}\;\;\;\Rightarrow\;\;\;\frac{kg^2_5}{R}=\frac{12\pi^2}{N}
\end{equation}
with $N$ the number of colors. Not surprisingly, this is also the
prediction of the Soft Wall model \cite{KKSS}, since it describes
the same UV physics as the Hard Wall model (the background dilaton
field, which smoothly warps the IR bulk region from $AdS_5$,
vanishes when $z\to0$).

The two-point correlator \eqref{HW vector QCD correlator} can also
be written as an infinite sum over the intermediate vector states
$|\rho_n^a(p,\l)\rangle$:
\begin{equation}\label{pole correlator}
\Pi_V^{(QCD)}(q^2)=-\sum_{n}\frac{F^2_{V_n}}{(q^2-m^2_{V_n}+i\ep)m^2_{V_n}}\;\;.
\end{equation}
The decay constants $F_{V_n}$ of the vector mesons are defined by
\begin{equation}
\langle0|j_{V\m}^a(x)|\rho_n^b(p,\lambda)\rangle=e^{-ip\cdot
x}\d^{ab}F_{V_n}\,\ve^{(\lambda)}_{\m}(p)
\end{equation}
where the polarization vectors $\ve^{(\lambda)}_{\m}(p)$ generate
(together with the momentum $p^{\m}$) a complete set in the sense
that
\begin{equation}
\sum_{\lambda}\ve^{(\lambda)}_{\m}(p)\ve^{(\lambda)}_{\n}(p)=-\eta_{\m\n}+\frac{p_{\m}p_{\n}}{p^2}
\end{equation}
with $\ve^{(\lambda)}(p)\cdot p=0$ and $\ve^{(\lambda)}(p)\cdot
\ve^{(\lambda')}(p)=-\d^{\lambda\lambda'}$. On the other hand,
Green's theorem applied to \eqref{HWM vector b-t-b} allows us to
express the vector bulk-to-boundary propagator in terms of a sum
over the vector resonances \cite{EKSS}:
\begin{equation}\label{IR HW Green Theorem vector}
\tilde{V}(q^2,z)=-\sum_n\frac{v_n(z)}{q^2-m_{V_n}^2+i\epsilon}\Big(\frac{1}{z'}\de_{z'}v_n(z')\Big)\Big|_{z'=\epsilon\to0}
\end{equation}
where the normalizable modes $v_n(z)$, solution of \eqref{HW
vector eq of motion} with $q^2=m_{V_n}^2$:
\begin{equation}\label{normalizable mode}
v_n(z)=\sqrt2\frac{z}{z_m}\frac{J_1(m_{V_n}z)}{J_1(m_{V_n}z_m)}\;\;,
\end{equation}
describes, according to the $AdS$/CFT dictionary, the vector
mesons and are subject to the following constraints:
\begin{eqnarray}
v_n(0)&=&0\;\;,\label{dirichlet boundary condition}\\
\de_zv_n(z)\Big|_{z=z_m}&=&0\;\;,\label{neumann boundary condition}\\
\int_0^{z_m}\frac{dz}{z}v_n(z)^2&=&1\;\;.\label{normalization}
\end{eqnarray}
Thanks to the IR Neumann boundary condition \eqref{neumann
boundary condition}, the Hard Wall model predicts a $\rho$ meson
mass spectrum: $m_{V_n}=\frac{\g_{0,n}}{z_m}$ given in terms of
the real zeros $\g_{0,n}$'s of the Bessel function of the first
kind $J_0(m_{V_n}z_m)=0$. One sets the value of $z_m$ from the
mass of the ground state $m_{\rho}\simeq 775$ MeV \cite{PDG}:
$z_m=\g_{0,1}/m_{\rho}\simeq1/323$ MeV$^{-1}\simeq0.61$ fm. From
the bulk-to-boundary propagator \eqref{IR HW Green Theorem
vector}, we have also the (square of the) vector meson decay
constants. Identifying \eqref{HW vector AdS correlation function}
with \eqref{pole correlator}, one finds \cite{EKSS}:
\begin{equation}\label{vector decay constant}
F_{V_n}^2=\frac{R}{kg_5^2}\Big(\frac{1}{z}\de_z
v_n(z)\Big)^2\Big|_{z=\epsilon\to0}=\Big(\frac{\g_{0,\,n}}{\sqrt{2}\pi\,z_m^2\,J_1(\g_{0,\,n})}\Big)^2\;\;.
\end{equation}
The case of the axial-vector current
$j_{A\m}^a(x)=\overline{q}\g_{\m}\g_5t^aq(x)$ is similar. The
two-point correlation function derived from large$-N$ QCD reads as
\cite{Pich}
\begin{equation}\label{App_two point AA correlator QCD}
\begin{array}{lll}
{\Pi^{(QCD)}_{A}}^{ab}_{\m\n}(q)&\equiv&i\int d^4x\,e^{iq\cdot
x}\langle0|T[j^a_{A\m}(x)j^b_{A\n}(0)]|0\rangle\\
\\
&=&\displaystyle\d^{ab}(q_{\m}q_{\n}-q^2\eta_{\m\n})\Big(-\frac{f_{\pi}^2}{q^2}-\sum_{n}\frac{F^2_{A_n}}{(q^2-m^2_{A_n}+i\ep)m^2_{A_n}}\Big)
\end{array}
\end{equation}
and has to be compared with
\begin{equation}\label{HW axial AdS correlation function}
{\Pi^{(AdS)}_{A}}^{ab}_{\m\n}(q)=\d^{ab}(q_{\m}q_{\n}-q^2\eta_{\m\n})\Big(\frac{1}{q^2}\frac{R}{kg_5^2}\tilde{A}(q^2,z)\frac{1}{z}\de_z\tilde{A}(q^2,z)\Big)\Big|_{z=\epsilon\to0}\;\;.
\end{equation}
The first term in \eqref{App_two point AA correlator QCD} is the
massless pion contribution. The decay constants of the
axial-vector mesons and of the pion, defined by:
\begin{equation}
\begin{array}{l}
\langle0|j^a_{A\m}(x)|a^b_{1\,n}(p,\lambda)\rangle=e^{-ip\cdot
x}\d^{ab}F_{a_{1\,n}}\ve^{(\lambda)}_{\m}(p)\;\;,\\
\langle0|j^a_{A\m}(x)|\pi^b(p)\rangle=e^{-ip\cdot
x}\d^{ab}if_{\pi}p_{\m}\;\;,
\end{array}
\end{equation}
take then the following $AdS$ forms \cite{EKSS}:
\begin{eqnarray}
F_{A_n}^2&=&\frac{R}{kg_5^2}\Big(\frac{1}{z}\de_z
a_n(z)\Big)^2\Big|_{z=\epsilon\to0}\;\;,\label{axial decay constant}\\
f_{\pi}^2&=&-\frac{R}{kg_5^2}\,\frac{1}{z}\de_z\tilde{A}(0,z)\Big|_{z=\epsilon\to0}\;\;.\label{pion
decay constant}
\end{eqnarray}
The axial-vector normalizable mode $a_n(z)$ is solution of
\eqref{HW b-to-b axial-vector EOM} with $q^2=m_{A_n}^2$ and
satisfies constraints similar to \eqref{dirichlet boundary
condition}-\eqref{normalization} while the decay constant of the
massless pion is given in terms of the axial-vector
bulk-to-boundary propagator $\tilde{A}(0,z)$, solution of
\eqref{HW axial-vector b-to-b} at $q^2=0$.

In section 3.3, we will be interested in the gravitational form
factor of the pion which enters the expression of the matrix
element of the traceless-transverse part $\hat{T}^{\m\n}$ of the
energy-momentum tensor between two one-pion states \cite{carlson}.
Following the $AdS$/CFT dictionary, to this operator corresponds,
in the dual theory, a $5d$ bulk field, namely the fluctuations
$h_{MN}(x,z)$ around the flat tensor metric in the so-called
Randall-Sundrum gauge:
\begin{equation}\label{h}
g_{MN}(x,z)=\frac{R^2}{z^2}\Big(\eta_{MN}+h_{MN}(x,z)\Big)
\end{equation}
with $h_{\mu z}=h_{zz}=0$, $\de^{\m}h_{\m\n}=0$ and
${h^{\m}}_{\m}=0$. In this gauge, the equations of motion derived
from \eqref{gravity part} take the simple form:
\begin{equation}
\de_z\big(\frac{1}{z^3}\de_z
\tilde{h}_{\m\n}(q,z)\big)+\frac{q^2}{z^3}\tilde{h}_{\m\n}(q,z)=0
\end{equation}
or
\begin{equation}\label{EOM gravity}
\Big[\de_z\big(\frac{1}{z^3}\de_z
\big)+\frac{q^2}{z^3}\Big]\tilde{h}(q^2,z)=0
\end{equation}
in terms of the tensor bulk-to-boundary propagator, defined as
usual through the relation \eqref{fourier convolution}:
\begin{equation}
\tilde{h}_{\m\n}(q,z)=\tilde{h}(q^2,z)\tilde{h}_{0\,\m\n}(q)
\end{equation}
and which satisfies the following UV and IR boundary conditions:
\begin{equation}
\tilde{h}(q^2,0)=1\;\;\;\;\textrm{and}\;\;\;\;\de_z\tilde{h}(q^2,z)\Big|_{z=z_m}=0\;\;.
\end{equation}
For space-like momentum $Q^2=-q^2>0$, the solution of \eqref{EOM
gravity} strongly resembles the expression \eqref{HW space-like
vector b-to-b} for the vector propagator ($Q\equiv\sqrt{Q^2}$):
\begin{equation}\label{tensor b-to-b propagator}
\tilde{h}(Q^2,z)=\frac{1}{2}Q^2z^2\Big(K_2(Qz)+\frac{K_1(Qz_m)}{I_1(Qz_m)}I_2(Qz)\Big)
\end{equation}
with the asymptotic behaviours:
\begin{eqnarray}
\tilde{h}(Q^2,z)&\underset{z\to0}{=}&1-\frac{Q^2z^2}{4}+O(z^4)\;\;,\\
\tilde{h}(Q^2,z)&\underset{Q^2\to0}{=}&1-\frac{Q^2z^2}{8}(2-\frac{z^2}{z_m^2})+O(Q^4)
\end{eqnarray}
such that $\tilde{h}(0,z)=1$.

\section{The $AdS$/QCD dictionary revised}

\subsection{The pion in the chiral limit}

In the chiral limit where the mass of the light quarks $m_q$ is
put to zero, the pion corresponds to a would-be Goldstone boson
associated with the spontaneous breaking of the chiral symmetry
and, as such, is massless. The Gell-Mann-Oakes-Renner (GMOR)
relation accounts for this behaviour: $m_{\pi}^2\propto m_q$. We
begin to introduce briefly the Hard Wall description of the pion
in the chiral limit \cite{GR}.

When $q^2=m_{\pi}^2=0$, the equations of motion \eqref{HW pion EOM
2} and \eqref{HW pion EOM 1} becomes (the subscript $\chi$ on the
pion wave functions refers to the chiral limit case):
\begin{equation}\label{HW chiral pion 1}
\de_z\Big(\frac{1}{z}\de_z\phi_{\chi}(z)\Big)-\frac{R^2 g_5^2
v(z)^2}{z^3}\Big(\phi_{\chi}(z)-\pi_{\chi}(z)\Big)=0
\end{equation}
and
\begin{equation}\label{HW chiral pion 2}
\de_z\pi_{\chi}(z)=0\;\;.
\end{equation}
The normalizable mode of the pion $\phi_{\chi}(z)$ satisfies the
UV and IR boundary conditions:
\begin{equation}
\phi_{\chi}(0)=0\;\;\;\;\textrm{and}\;\;\;\;\de_z\phi(z)\Big|_{z=z_m}=0
\end{equation}
while $\pi_{\chi}(z)$ is a constant of $z$. Then, it is convenient
to define a new bulk field, the so-called chiral field, as
$\Psi(z)\equiv\phi_{\chi}(z)-\pi_{\chi}(z)$ which is solution to
\eqref{HW chiral pion 1}:
\begin{equation}\label{chiral fiel EOM}
\de_z\Big(\frac{1}{z}\de_z\Psi(z)\Big)-\frac{R^2 g_5^2
v(z)^2}{z^3}\Psi(z)=0\;\;.
\end{equation}
Moreover, if $\pi_{\chi}(z)=-1$ then $\Psi(z)$ verifies the
constraints:
\begin{equation}\label{bc chiral field}
\Psi(0)=1\;\;\;\;\textrm{and}\;\;\;\;\de_z\Psi(z)\Big|_{z=z_m}=0\;\;.
\end{equation}
It turns out that the equation of motion \eqref{chiral fiel EOM}
as well as the boundary conditions \eqref{bc chiral field} are
also those of the the axial-vector bulk-to-boundary propagator
$\tilde{A}(0,z)$ evaluated at $q^2=0$ (see Eqs.\eqref{HW
axial-vector b-to-b}-\eqref{HWM IR bc}). As a result, we have the
identify:
\begin{equation}
\Psi(z)=\tilde{A}(0,z)\;\;.
\end{equation}

On the other hand, the background field $v(z)$ in \eqref{chiral
fiel EOM} is solution of \eqref{HW vev EOM} and reads
\begin{equation}\label{v}
v(z)=\frac{\overline{m}}{R}z+\frac{\overline{\sigma}}{R}z^3
\end{equation}
where two parameters $\overline{m}$ and $\overline{\sigma}$, which
do not have to be identified \emph{a priori} with the quark mass
$m_q$ and the chiral condensate
$\sigma\equiv-\langle\overline{q}q\rangle$, have been introduced
\cite{hard wall model correlator,CCW,Gherghetta,krikun2}.
Nevertheless, since the two contributions of $v(z)$ on the
\emph{r.h.s.} of \eqref{v} accounts for the explicit and implicit
chiral symmetry breaking respectively, it is natural to assume a
proportionality relation between, on the one hand, $\overline{m}$
and $m_q$ and, on the other hand, $\overline{\sigma}$ and
$\sigma$. Thus, in the chiral limit we are interested in, only
remains the second term:
\begin{equation}
v(z)=\frac{\overline{\sigma}}{R}z^3
\end{equation}
and \eqref{chiral fiel EOM} becomes (one defines
$\alpha\equiv\frac{g_5\overline{\sigma}}{3}$):
\begin{equation}
\de_z\Big(\frac{1}{z}\de_z\Psi(z)\Big)-9\alpha^2z^3\Psi(z)=0
\end{equation}
with the solution \cite{GR}:
\begin{equation}\label{chiral field first expression}
\Psi(z)=\G\big(\frac{2}{3}\big)\Big(\frac{\a}{2}\Big)^{1/3}\,z\,\Big[I_{-1/3}(\a
z^3)-\frac{I_{2/3}(\a z_m^3)}{I_{-2/3}(\a z_m^3)}I_{1/3}(\a
z^3)\Big]\;\;.
\end{equation}
The decay constant of the massless pion \eqref{pion decay
constant} takes then the form:
\begin{equation}\label{pion decay constant 2}
f_{\pi}^2=-\frac{R}{kg^2_5}\frac{1}{z}\de_z\Psi(z)\Big|_{z=\epsilon\to0}=\frac{R}{kg_5^2}\,3\cdot
2^{1/3}\frac{\G(2/3)}{\G(1/3)}\frac{I_{2/3}(\alpha
z_m^3)}{I_{-2/3}(\alpha z_m^3)}\alpha^{2/3}\;\;.
\end{equation}

\subsection{The chiral symmetry breaking function and the $AdS$/QCD dictionary}

The background field $v(z)$ \eqref{v} involves two parameters
$\overline{m}$ and $\overline{\sigma}$ which remains to identify.
As they account for the chiral symmetry breaking, they must be
related, in some way, to the light quark mass $m_q$ and the chiral
condensate $\sigma$. The latter enter the GMOR relation which
relates the magnitude of the pion and quark masses to the size of
the quark condensate:
\begin{equation}\label{GMOR}
m_{\pi}^2f_{\pi}^2=2m_q\sigma+O(m_{q}^2)\;\;.
\end{equation}
In the chiral limit, $m_q=0$ and the pion is massless. It becomes
massive provided the chiral symmetry is explicitly broken.
Although the result is well-known and consists of one of the first
insights in support of the Hard Wall model, it is worth summing up
shortly the main steps leading to \eqref{GMOR} as performed
initially by \cite{EKSS}. When $q^2=m_{\pi}^2$, the pseudo-scalar
normalizable modes $\phi(z)$ and $\pi(z)$ are not independent but
satisfy two coupled differential equations \eqref{HW pion EOM 2}
and \eqref{HW pion EOM 1}. In particular, we have:
\begin{equation}\label{normalizable mode relation}
m_{\pi}^2\de_z\phi(z)=\frac{R^2g^2_5v(z)^2}{z^2}\de_z\pi(z)\;\;.
\end{equation}
Moreover, they behave asymptotically as
\begin{equation}
\phi(0)=0\;\;\;\;\textrm{and}\;\;\;\;\de_z\phi(z)\Big|_{z_m}=0\;\;,
\end{equation}
\begin{equation}\label{UV pion bc}
\pi(0)=0\;\;.
\end{equation}
The IR boundary condition of $\pi(z)$ is, on the other hand,
dictated by the chiral symmetry: since $\pi_{\chi}(z)=-1$ in the
chiral limit, the value of the pion wave function $\pi(z)$ must
approach $-1$ at low energies, \emph{i.e.} near the IR cutoff
$z_m$. Let us rewrite \eqref{normalizable mode relation} as
follows:
\begin{equation}
\pi(z)=m_{\pi}^2\int_0^z du\frac{u^3}{R^2v(u)^2}\frac{1}{g_5^2
u}\de_u\phi(u)
\end{equation}
where the lower bound of the integral is given by the UV boundary
condition \eqref{UV pion bc}. Since there is a mass gap separating
the light pion from the rest of the hadronic spectrum, it is
consistent to expand the pion wave functions in powers of
$m_{\pi}^2$ in such a way that
\begin{equation}\label{GMOR2}
\pi(z)=m_{\pi}^2\int_0^z du f(u)\frac{1}{g_5^2 u}\de_u\Psi(u)
\end{equation}
at leading order. The function $f(u)=\frac{u^3}{R^2v(u)^2}$
exhibits a peak located at
$u_c=\sqrt{\frac{\overline{m}}{3\overline{\sigma}}}$. The main
contribution to the integral comes from the neighbourhood of
$u_c$. Provided that $\overline{m}$ is small in comparison with
$\overline{\sigma}$, which is consistent with the fact that
$\overline{m}$ and $\overline{\sigma}$ are related to $m_q$ and
$\sigma$ respectively, only small values of $u$ give sizeable
contributions to the integral. Consequently, the remaining
function $\frac{1}{g_5^2 u}\de_u\Psi(u)$ can be evaluated at
$u\simeq0$ which is then related to the pion decay constant
\eqref{pion decay constant 2}:
\begin{equation}
\pi(z)=-\frac{k}{R}m_{\pi}^2 f_{\pi}^2\int_0^z du f(u)\;\;.
\end{equation}
It remains to perform the integration. When $z$ is large, $\pi(z)$
approaches -1 so that
\begin{equation}
-1=-\frac{k}{R}\frac{m_{\pi}^2
f_{\pi}^2}{2\overline{m}\,\overline{\sigma}}\;\;.
\end{equation}
In order to recover the GMOR relation at leading order
\eqref{GMOR}, we have thus to impose the simple condition:
\begin{equation}\label{condition}
\overline{m}\,\overline{\sigma}=\frac{k}{R}m_q\sigma\;\;.
\end{equation}
It is then consistent to take:
\begin{eqnarray}
\overline{m}&=&m_q\;\;,\\
\overline{\sigma}&=&\frac{k}{R}\sigma\;\;.
\end{eqnarray}

The $AdS$/QCD matching at short distances of the two-point
correlation function of the vector current \eqref{HW overall
constant vector meson} allowed us to extract the factor
$\frac{kg_5^2}{R}=\frac{12\pi^2}{N}$ \cite{EKSS}. This value is
not peculiar to the Hard Wall model. Any holographic model must
indeed be conformal near the UV brane and, as such, must predict
the same UV physics. In particular, it is not surprising to
recover this value in the Soft Wall model \cite{KKSS}: the
background dilaton field $\Phi(z)$ vanishes when $z\to0$ while the
bulk reduces asymptotically to $AdS_5$. On the other hand, light
scalar mesons have also been studied in the framework of the Soft
Wall model \cite{soft wall scalar} where the $5d$ effective action
consists in coupling precisely the Lagrangian \eqref{IR hard wall
action} (with the mostly plus signature) to the background
dilation field through the overall factor $e^{-\Phi(z)}$. In this
framework, the scalar bulk field $S^A(x,z)$, dual to the QCD
operator $\mathcal{O}^A_S(x)=\overline{q}(x)T^Aq(x)$, includes in
\eqref{scalar bulk field} singlet $S_1(x,z)$ and non-singlet
$S^a(x,z)$ components, gathered into the multiplet
$S=S^AT^A=S_1T^0+S^aT^a$ with $T^0=\mathbb{I}_{n_F}/\sqrt{2n_F}$
and $T^a$ the generators of $SU(n_F)$, with normalization
$Tr\Big(T^AT^B\Big)=\d^{AB}/2$ ($A=0,a$ and $a=1,\ldots,n_F^2-1$).
The mass spectrum and the decay constants of the scalar mesons
have been determined as the poles and the residues of the
two-point scalar current correlation function, defined in QCD as
\begin{equation}\label{QCD scalar correlator}
\Pi^{(QCD)AB}_S(q^2)\equiv i\int d^4x e^{iq\cdot
x}\langle0|T[\mathcal{O}^A(x)\mathcal{O}^B(0)]|0\rangle
\end{equation}
and which involves in $AdS$ side the scalar bulk-to-boundary
propagator \cite{soft wall scalar}. We have:
\begin{equation}\label{AdS scalar correlator}
\Pi^{(AdS)AB}_S(q^2)=\d^{AB}\frac{R^3}{k}\tilde{S}(\frac{q^2}{c^2},c^2z^2)\frac{e^{-\Phi(z)}}{z^3}\de_{z}\tilde{S}(\frac{q^2}{c^2},c^2z^2)\Big|_{z=\epsilon\to0}
\end{equation}
where
\begin{equation}\label{scalar b-to-b propagator}
\tilde{S}(\frac{q^2}{c^2},c^2z^2)=\frac{1}{Rc}\Gamma(\frac{q^2}{4c^2}+\frac{3}{2})(cz)U(\frac{q^2}{4c^2}+\frac{1}{2};0;c^2z^2)\underset{z\to0}{=}\frac{z}{R}\;\;.
\end{equation}
The dilaton parameter $c$ breaks the conformal invariance of the
boundary field theory (\emph{i.e.} breaks the isometry group of
$AdS_5$) and gives the dimension for all the observables as the
masses, the decay constants, the condensates and so on and so
forth.

Especially, the $AdS$ expression of the correlator can be
compared, at high energies, with its computation in QCD (for
$N=3$), which enables us to extract the factor $\frac{k}{R}$:
\begin{equation}\label{SW scalar operator correlator}
\left.
\begin{array}{l}
\Pi^{(AdS,pert.)AB}_S(Q^2)=\d^{AB}\frac{R}{k}Q^2\ln(Q^2z^2)\Big|_{z=\frac{1}{\n}}\\
\Pi^{(QCD,pert.)AB}_S(Q^2)=\frac{\d^{AB}}{2}\frac{3}{8\pi^2}Q^2\ln(\frac{Q^2}{\n^2})
\end{array}
\right\}\;\;\;\Rightarrow\;\;\;\frac{k}{R}=\frac{16\pi^2}{N}
\end{equation}
and, subsequently, $g_5^2=3/4$ according to \eqref{HW overall
constant vector meson}. Being associated with the UV physics at
small $z$, the value \eqref{SW scalar operator correlator} can
also be used in the framework of the Hard Wall model since we are
considering two \emph{equivalent} actions, regardless of the
dilaton overall factor (Compare the action \eqref{IR hard wall
action} with Eq.(3) of \cite{soft wall scalar}) and, consequently,
the same UV boundary condition \eqref{UV boundary condition of v}.
As a result, the chiral symmetry breaking function \eqref{v} takes
the following form:
\begin{equation}\label{v1}
v(z)=\frac{z}{R}(m_q+\frac{16\pi^2}{N}\sigma
z^2)\underset{z\to0}{=}\frac{m_q}{R}z\;\;.
\end{equation}
Its large$-N$ scaling is consistent since $m_q\sim O(N^0)$ and
$\sigma\sim O(N)$ yield $v(z)\sim O(N^0)$.

Let us now clarify the $AdS$/QCD dictionary. As in \cite{CCW}, we
start by considering the singlet case $A=B=0$ of \eqref{AdS scalar
correlator} and \eqref{SW scalar operator correlator} with the
relation: $\frac{2R}{k}=\frac{3}{8\pi^2}=\frac{a^2}{2}$, where the
parameter $a=\frac{\sqrt{N}}{2\pi}$ has been introduced in
\cite{CCW}. We then perform a mere rescaling of the scalar bulk
field $X(x,z)$ and of the $5d$ coupling constant:
\begin{equation}\label{redefinition}
\hat{X}(x,z)=\frac{a}{2}\,X(x,z)\;\;\;\textrm{and}\;\;\;\frac{1}{\hat{g}_5^2}=\frac{a^2}{4}\,\frac{1}{g_5^2}\;\;.
\end{equation}
That implies the following renormalization of the $5d$ holographic
action \eqref{IR hard wall action}:
\begin{equation}\label{renormalization}
\frac{1}{k}\int
d^5x\sqrt{g}\,Tr\,\mathcal{L}(X,g_5^2,\pi,A_L,A_R)\;\;\;\Rightarrow\;\;\;\frac{1}{R}\int
d^5x\sqrt{g}\,Tr\,\mathcal{\hat{L}}(\hat{X},\hat{g}_5^2,\pi,A_L,A_R)\;\;.
\end{equation}
The action $S_{5d}^{(matter)}$, written in the first form,
involves an overall factor $1/k$ related to the number of colors
$N$ \cite{pomarol,hard wall model correlator}. Because of the
scalar bulk field and gauge coupling redefinition
\eqref{redefinition}, this $N$-dependent factor is absorbed on the
\emph{r.h.s.} of \eqref{renormalization}. That implies
modifications in the field/operator prescription. The chiral
symmetry breaking function \eqref{v1} becomes indeed:
\begin{equation}\label{v2}
\hat{v}(z)=\frac{a}{2}\,\frac{m_q}{R}z+\frac{2}{a}\,\frac{\sigma}{R}z^3\underset{z\to0}{=}\frac{a}{2}\frac{m_q}{R}z\;\;.
\end{equation}
According to \eqref{v}, we have $\overline{m}=\frac{a}{2}\,m_q$
and $\overline{\sigma}=\frac{2}{a}\,\sigma$ which fulfils the
condition $\overline{m}\,\overline{\sigma}=m_q\sigma$, analogous
to \eqref{condition}. Moreover, the large$-N$ behaviour has
changed, being now $\hat{v}(z)\sim O(\sqrt{N})$. Besides, one
usually chooses to renormalize the $5d$ gauge coupling such that
$\hat{g}_5^2=\frac{12\pi^2}{N}$ instead of the gauge bulk fields
$A_{L_M}^a$ and $A_{R_M}^a$ (or, equivalently, $V^a_{M}$ and
$A^a_M$), which has the advantage of letting unchanged the UV
boundary conditions \eqref{HWM UV bc}. Also, the pseudo-scalar
modes $\pi$ and $\phi$ are not affected by the rescaling
\eqref{redefinition}. As a result, the equations of motion
\eqref{HW vev EOM}-\eqref{HW pion EOM 1} and \eqref{chiral fiel
EOM} and the boundary conditions \eqref{bc chiral field} remain
the same with $g_5^2$ and $v(z)$ replaced by $\hat{g}_5^2$ and
$\hat{v}(z)$ (let us point out that the products $g_5^2\,v(z)^2$
and $\hat{g}_5^2\,\hat{v}(z)^2$ are both $N$-independent). As for
the decay constants \eqref{vector decay constant}, \eqref{axial
decay constant} and \eqref{pion decay constant}, they retain their
$AdS$ expressions except for the overall factor $\frac{R}{k
g_5^2}$ which becomes $\frac{1}{\hat{g}_5^2}$. Finally, the scalar
bulk-to-boundary propagator \eqref{scalar b-to-b propagator} gains
an additional factor of $\frac{a}{2}$ so as to recover
$a=\frac{\sqrt{N}}{2\pi}$ from the matching \eqref{SW scalar
operator correlator}.

The key point in the discussion above referred to the different
operators at work: while the source of the complex scalar bulk
field $X(x,z)$ is associated with the operator
$\overline{q}_Rq_L$, to the background field $v(z)$ and, for the
sake of clearness, to the singlet component $S_1(x,z)$ of the real
scalar bulk field $S(x,z)=S^A(x,z)T^A$ correspond, in the isospin
limit, the vacuum expectation value of
$\overline{q}q=2\,\textit{Re}(\overline{q}_Rq_L)$ and the operator
$\sqrt{2n_F}\,\textit{Re}(\overline{q}_Rq_L)$ respectively. Hence,
we can see the relevance of the factors of $1/2$ and of
$1/\sqrt{2n_F}$ in front of $v(z)$ and of $S_1$ in \eqref{scalar
bulk field} in order to have the UV asymptotic behaviours
\eqref{UV boundary condition of v} and \eqref{scalar b-to-b
propagator}.

\subsection{The large$-N$ behaviour of the Hard Wall model of QCD}

In the preceding section, we have determined unambiguously the
light quark mass and the chiral condensate entering the expression
of the chiral symmetry breaking function \eqref{v1} which enabled
us to clarify the $AdS$/QCD dictionary as used in the
\emph{bottom-up} approach of the holographic models of QCD. We
return to our main goal focusing on the large$-N$ behaviour of the
Hard Wall model. Our starting point was the $5d$ effective action
\eqref{action}.  A glance at the bulk fields leads us to conclude
that they do not scale with $N$ (see, for instance, the
normalizable modes \eqref{normalizable mode} and \eqref{chiral
field first expression}, the bulk-to-boundary propagators
\eqref{HW space-like vector b-to-b} and \eqref{tensor b-to-b
propagator}, as well as the chiral symmetry breaking function
\eqref{v1}). As a consequence, the square of the decay constants
$F_{V_n}^2$, $F_{A_n}^2$ and $f_{\pi}^2$ of the vector,
axial-vector and pseudo-scalar mesons, respectively \eqref{vector
decay constant}, \eqref{axial decay constant} and \eqref{pion
decay constant}, scale as $O(N)$. This is reminiscent of what
happens in Chiral Perturbation Theory with the large$-N$ behaviour
implemented \cite{Pich}. Although already conclusive, let us go
one step further in our check by considering the electromagnetic
\cite{GR} and gravitational \cite{carlson} form factors of the
pion in the chiral limit.

The vector form factor $F_{\pi}(q^2)$ of the pion can be obtained
from the three-point correlation function:
\begin{equation}
{\Pi^{(QCD)}}^{abc}_{\alpha\mu\beta}(p_1,p_2)\equiv-\int
d^4xd^4y\,e^{ip_1\cdot x-ip_2\cdot
x}\langle0|T[j^a_{A\alpha}(x)j^b_{V\m}(0)j^c_{A\b}(y)]|0\rangle
\end{equation}
where the vector current $j^b_{V\m}$ is sandwiched between the two
axial-vector currents $j^a_{A\alpha}$ and $j^c_{A\b}$ which
project on a one-pion state. So, according to the duality relation
\eqref{generating}, we need to consider in $AdS$ side the
following interaction terms \cite{GR,KL}:
\begin{equation}\label{interaction term vectorf.f.}
S_{5d}^{(VPP)}=\frac{\epsilon^{abc}}{k}\int
d^5x\frac{R}{z}\Big[\frac{1}{g_5^2}\Big(\de_z\de^{\mu}\phi^a\Big)V_{\m}^b\Big(\de_z\phi^c\Big)+\frac{R^2v(z)^2}{z^2}\Big(\de^{\m}\phi^a-\de^{\m}\pi^a\Big)V_{\m}^b\Big(\phi^c-\pi^c\Big)\Big]\;\;,
\end{equation}
the derivation of it being somewhat tedious as involving the use
of the equations of motion \eqref{HW vector eq of motion} and
\eqref{HW pion EOM 2} as well as several partial integrations. In
the $4d$ Fourier space, one gets:
\begin{eqnarray}
S_{5d}^{(VPP)}&=&\int\frac{d^4q_1d^4q_2d^4q_3}{(2\pi)^{12}}\,i(2\pi)^4\d^4(q_1+q_2+q_3)q_1^{\m}\times\non\\
&&\frac{\epsilon^{abc}}{k}\int_0^{z_m}\frac{R}{z}\tilde{V}^b_{\m}(q_2,z)\Big[\frac{1}{g_5^2}\Big(\de_z\tilde{\psi}^a(q_1,z)\Big)\Big(\de_z\tilde{\psi}^c(q_3,z)\Big)+\frac{R^2v(z)^2}{z^2}\tilde{\psi}^a(q_1,z)\tilde{\psi}^c(q_3,z)\Big]\non\\
\end{eqnarray}
where
\begin{equation}\label{goldstone bulk field}
\tilde{\psi}^a(q_1,z)=\frac{1}{q_1^2}\Psi(z)\Big(-iq_1^{\alpha}\tilde{A}^a_{\parallel0\alpha}(q_1)\Big)
\end{equation}
is the Goldstone boson bulk field expressed in terms of the chiral
field $\Psi(z)$ and the longitudinal component
$\tilde{A}^a_{\parallel0\alpha}(q_1)$ of the axial-vector source
field. Since it describes a massless mode, only the pole $q_1^2=0$
appears in \eqref{goldstone bulk field}. Then, three functional
derivatives with respect to the vector $\tilde{V}^b_{0\,\m}(q)$
and the axial-vector $\tilde{A}^a_{\parallel0\alpha}(p_1)$ and
$\tilde{A}^c_{\parallel0\beta}(-p_2)$ sources gives the $AdS$
expression of the three-point correlation function ($q=p_2-p_1$):
\begin{equation}
{\Pi^{(AdS)}}^{abc}_{\alpha\m\beta}(p_1,p_2)=\frac{p_{1\alpha}p_{2\alpha}}{p_1^2p_2^2}(p_1+p_2)_{\m}f_{\pi}^2F_{\pi}(q^2)\;\;.
\end{equation}
Hence, the electromagnetic form factor of the pion which, for
space-like momentum $Q^2=-q^2<0$, reads as
\begin{equation}\label{pion form factor chiral limit}
F_{\pi}(Q^2)=\frac{R}{kg_5^2}\frac{1}{f_{\pi}^2}\int_0^{z_m}dz\,z\,
\tilde{V}(Q^2,z)\Big[\Big(\frac{\de_z\Psi(z)}{z}\Big)^2+\frac{R^2g_5^2v(z)^2}{z^4}\Psi(z)^2\Big]
\end{equation}
where $\tilde{V}(Q^2,z)$ is the vector bulk-to-boundary propagator
\eqref{HW space-like vector b-to-b} and which satisfies the charge
normalization $F_{\pi}(0)=1$ at $Q^2=0$. The property
$\tilde{V}(0,z)=1$ and the equation of motion for $\Psi(z)$
\eqref{chiral fiel EOM} allow indeed to write out successively:
\begin{eqnarray}
F_{\pi}(0)&=&\frac{R}{kg_5^2}\frac{1}{f_{\pi}^2}\int_{0}^{z_m}dz\,z\Big[\Big(\frac{\de_{z}\Psi(z)}{z}\Big)^2+\frac{R^2g_5^2v(z)^2}{z^4}\Psi(z)^2\Big]\\
&=&\frac{R}{kg_5^2}\frac{1}{f_{\pi}^2}\int_{0}^{z_m}dz\,z\Big[\Big(\frac{\de_z\Psi(z)}{z}\Big)^2+\Psi(z)\frac{1}{z}\de_z\Big(\frac{\de_z\Psi(z)}{z}\Big)\Big]\\
F_{\pi}(0)&=&\frac{R}{kg_5^2}\frac{1}{f_{\pi}^2}\int_{0}^{z_m}dz\,\de_z\Big(\Psi(z)\frac{1}{z}\de_z\Psi(z)\Big)\;\;.
\end{eqnarray}
The integration at $z_m$ gives zero because of the IR Neumann
boundary condition \eqref{bc chiral field}. Thus, one obtains:
\begin{equation}
F_{\pi}(0)=\frac{1}{f_{\pi}^2}\Psi(z)\Big(-\frac{R}{kg_5^2}\frac{1}{z}\de_z\Psi(z)\Big)\Big|_{z=\epsilon\to0}=1
\end{equation}
thanks to the UV Dirichlet boundary condition \eqref{bc chiral
field} and to the $AdS$ expression of the massless pion decay
constant \eqref{pion decay constant 2}. As it must be, the vector
form factor does not scale with $N$ which stems from the
compensation mechanism between $\frac{R}{kg_5^2}\sim O(N)$ and
$1/f_{\pi}^2\sim O(1/N)$ \cite{Pich}. As for the strong $VPP$
coupling constants, they follow writing the vector
bulk-to-boundary propagator in terms of the vector mass poles
$m_{V_n}$, of the residues $F_{V_n}$ \eqref{vector decay constant}
and of the normalizable eigenfunction $v_n(z)$ \eqref{normalizable
mode}. Eq. \eqref{IR HW Green Theorem vector} gives:
\begin{equation}
\tilde{V}(q^2,z)=-\sqrt{\frac{kg_5^2}{R}}\sum_{n=1}^{\infty}\frac{F_{V_n}\,v_n(z)}{q^2-m_{V_n}^2+i\epsilon}\;\;.
\end{equation}
(On the way, one verifies that the propagator does not scale with
$N$: $\tilde{V}(q^2,z)\sim O(N^0)$.) The $AdS$ expression of the
vector form factor and of the the $g_{V_n\pi\pi}$ couplings
follow:
\begin{equation}
F_{\pi}(q^2)=-\sum_{n=1}^{\infty}\frac{F_{V_n}\,g_{V_n\pi\pi}}{q^2-m_n^2+i\epsilon}
\end{equation}
with
\begin{equation}
g_{V_n\pi\pi}=\sqrt{\frac{R}{kg_5^2}}\frac{1}{f_{\pi}^2}\int_0^{z_m}dz\,z\,v_n(z)\,\Big[\Big(\frac{\de_z\Psi(z)}{z}\Big)^2+\frac{R^2g_5^2v(z)^2}{z^4}\Psi(z)^2\Big]\;\;.
\end{equation}
As a result, the large$-N$ counting rules are consistent in the
$AdS$/QCD Hard Wall model as $g_{V_n\pi\pi}\sim O(1/\sqrt{N})$
such that we recover free, stable, non-interacting mesons at
$N\to\infty$.

As a matter of fact, the study of the gravitational from factor of
the pion in the chiral limit gives very similar formulae
\cite{carlson}. Since the longitudinal part of the axial-vector
bulk field becomes physical and describes pions provided that the
chiral symmetry is broken, the interaction terms must involve all
the couplings of the type $h\pi\pi$, $hA\pi$ and $hAA$ where
$h_{\m\n}$ stands for perturbations of the flat tensor metric
\eqref{h}. One has:
\begin{equation}\label{grav interaction terms}
S^{(TPP)}=\frac{1}{k}\int
d^5x\frac{R}{z}h^{\m\n}(x,z)\Big[-\frac{R^2v(z)^2}{2z^2}\big(\de_{\m}\pi^a-A^a_{\m}\big)\big(\de_{\n}\pi^a-A^a_{\n}\big)
+\frac{1}{2g_5^2}\big(-F^a_{A_{\m z}}F^a_{A_{\n
z}}+\eta^{\alpha\beta}F^a_{A_{\n\a}}F^a_{A_{\m\b}}\big)\Big]\;\;.
\end{equation}
As usual, the functional derivatives with respect to the source
fields $\tilde{h}_{0\,\m\n}(q)$,
$\tilde{A}^a_{\parallel0\,\alpha}(p_1)$ and
$\tilde{A}^c_{\parallel0\,\beta}(-p_2)$ gives the three-point
correlation function including the transverse-traceless part
$\hat{T}^{\m\n}$ of the energy-momentum tensor and from which can
be derived the gravitational form factor of the pion\footnote{The
transverse-traceless part of the stress tensor is selected by
substituting in (the $4d$ Fourier transform of) \eqref{grav
interaction terms} $\tilde{h}_0^{\m\n}(q)$ with
$\Big[\Big(\eta^{\m\rho}-\frac{q^{\m}q^{\rho}}{q^2}\Big)\Big(\eta^{\n\sigma}-\frac{q^{\n}q^{\sigma}}{q^2}\Big)-\frac{1}{3}\Big(\eta^{\m\n}-\frac{q^{\m}q^{\n}}{q^2}\Big)\Big(\eta^{\rho\sigma}-\frac{q^{\rho}q^{\sigma}}{q^2}\Big)\Big]\tilde{h}_{0\,\rho\sigma}(q)$.}:
\begin{equation}\label{gravitational form factor of the pion}
A_{\pi}(Q^2)=\frac{R}{kg_5^2}\frac{1}{f_{\pi}^2}\int_0^{z_m}dz\,z
\,\tilde{h}(Q^2,z)\Big[\Big(\frac{\de_z\Psi(z)}{z}\Big)^2+\frac{R^2g_5^2v(z)^2}{z^4}\Psi(z)^2\Big]\;\;.
\end{equation}
This expression is quite similar to the electromagnetic form
factor \eqref{pion form factor chiral limit} where the vector
bulk-to-boundary propagator \eqref{HW space-like vector b-to-b} is
replaced by the tensor one \eqref{tensor b-to-b propagator}. Since
$\tilde{h}(0,z)=1$, the normalization $A_{\pi}(0)=1$ follows and
one checks the large$-N$ behaviour for the form factor:
$A_{\pi}(Q^2)\sim O(N^0)$.

For the sake of completeness, let us conclude this note by
performing some numerical estimates. For this, it is convenient to
introduce the so-called radial distribution density \cite{GR}:
\begin{equation}
\rho(\hat{z},c)=\frac{R}{kg_5^2}\frac{1}{f_{\pi}^2\,z_m^2}\Big[\Big(\frac{\de_{\hat{z}}\Psi(\hat{z})}{\hat{z}}\Big)^2+9\,c^2\hat{z}^2\Psi(\hat{z})^2\Big]
\end{equation}
with
\begin{equation}\label{chiral field first expression2}
\Psi(\hat{z})=\G\big(\frac{2}{3}\big)\Big(\frac{\beta}{2}\Big)^{1/3}\,\hat{z}\,\Big[I_{-1/3}(\beta\hat{z}^3)-\frac{I_{2/3}(\beta)}{I_{-2/3}(\beta)}I_{1/3}(\beta\hat{z}^3)\Big]\;\;.
\end{equation}
$\hat{z}=z/z_m$ is the dimensionless holographic coordinate while
$\beta=\alpha z_m^3$ and $\alpha=\frac{g_5}{3}\frac{k}{R}\sigma$.
Imposing that the pion decay constant \eqref{pion decay constant
2} coincides with the value $f_{\pi}\simeq92.4$ MeV in the chiral
limit, one gets $\alpha\simeq(424\;\textrm{MeV})^3$ and
$\beta\simeq2.27$ with $1/z_m\simeq323$ MeV determined from the
mass of the vector ground state. Because of the factor of
$\frac{k}{R}$ in the definition of $\alpha$, the quark condensate
$\sigma\simeq(171\;\textrm{MeV})^3$ turns out to be noticeably
smaller than Hard Wall model estimates where this factor is
usually omitted. On the other hand, the pion form factors
\eqref{pion form factor chiral limit} and \eqref{gravitational
form factor of the pion} become:
\begin{eqnarray}
F_{\pi}(Q^2)&=&\int_0^1
d\hat{z}\,\hat{z}\,\tilde{V}(Q^2,\hat{z})\rho(\hat{z},c)\;\;,\label{EM form factor}\\
A_{\pi}(Q^2)&=&\int_0^1
d\hat{z}\,\hat{z}\,\tilde{h}(Q^2,\hat{z})\rho(\hat{z},c)\label{GRAV
form factor}
\end{eqnarray}
from which can be obtained the pion radii defined as the slope at
$Q^2=0$ of the corresponding form factors \cite{GR,carlson}:
\begin{eqnarray}
\langle r_{\pi}^2\rangle_C&\equiv&-6\frac{d F_{\pi}(Q^2)}{d
Q^2}\Big|_{Q^2=0}=\frac{3}{2}z_m^2\int_0^1
d\hat{z}\,\hat{z}^3\,\Big(1-2\ln(\hat{z})\Big)\rho(\hat{z})\simeq(0.57\;\textrm{fm})^2\;\;,\non\\
\label{HW pion charge radius}\\
\langle
r_{\pi}^2\rangle_{grav.}&\equiv&-6\frac{dA_{\pi}(Q^2)}{dQ^2}\Big|_{Q^2=0}=\frac{3}{2}z_m^2\int_0^1d\hat{z}\,\hat{z}^3(1-\frac{\hat{z}^2}{2})\rho(\hat{z})\simeq(0.36\;\textrm{fm})^2\;\;,
\end{eqnarray}
to be compared with the experimental values $\langle
r_{\pi}^2\rangle_C\simeq(0.67\;\textrm{fm})^2$ \cite{PDG}. On the
other hand, the charge radius turns out to be larger than the
gravitational size which means that the Hard Wall model predicts a
more compactified pion momentum density distribution than for its
charge density \cite{carlson}.

\section{Conclusion}

This note aimed at clarifying some confusion regarding the
large$-N$ 't Hooft limit of holographic models of QCD. It is in
line with an approach consisting in reducing as much as possible
the ambiguities stemming from the modifications of the $AdS$/CFT
correspondence towards any non-$AdS$/non-CFT duality. In
particular, the devising concerned the large$-N$ behaviour of the
Hard-Wall model which, used as a guideline, led us to study the
field/operator prescription. Such subtleties are not
systematically taken into account when dealing with different
$AdS$/QCD models.

\end{document}